\documentclass[twocolumn,preprintnumbers,showpacs]{revtex4}
\usepackage{amsmath}
\usepackage{amssymb}
\usepackage{graphicx}
\usepackage{color}

\newcommand{\BAN}{\ensuremath{B_{1g}\,}}
\newcommand{\BN}{\ensuremath{B_{2g}\,}}

\begin{document}

\title{Collapse of the Normal State Pseudogap at a Lifshitz Transition in Bi$_{2}$Sr$_{2}$CaCu$_{2}$O$_{8+\delta}$
Cuprate Superconductor}

\author{S. Benhabib$^1$, A. Sacuto$^1$, M. Civelli$^2$, I. Paul$^1$, M. Cazayous$^1$, Y. Gallais$^1$,
M.-A. M\'easson$^1$, R. D. Zhong$^3$, J. Schneeloch$^3$ and G. D. Gu$^3$, D. Colson$^4$ and A. Forget$^4$}

\affiliation{$^1$ Laboratoire Mat\'eriaux et Ph\'enom$\grave{e}$nes Quantiques (UMR 7162 CNRS),
Universit\'e Paris Diderot-Paris 7, Bat. Condorcet, 75205 Paris Cedex 13, France,\\
$^2$ Laboratoire de Physique des Solides,(UMR 8502 CNRS), Universit\'e Paris Sud, Bat.510, 91405 Orsay Cedex,\\
$^3$ Matter Physics and Materials Science, Brookhaven National Laboratory (BNL), Upton, NY 11973, USA,\\
$^4$Service de Physique de l'Etat Condens\'{e}, CEA-Saclay, 91191 Gif-sur-Yvette, France}

\date{\today}

\begin{abstract}

We report a fine tuned doping study of strongly overdoped Bi$_2$Sr$_2$CaCu$_2$O$_{8+\delta}$ single crystals using
electronic Raman scattering.  Combined with theoretical calculations, we show that the doping, at which the normal state
pseudogap closes, coincides with a Lifshitz quantum phase transition where the active hole-like Fermi surface becomes
electron-like. This conclusion suggests that the microscopic cause of the
pseudogap is sensitive to the Fermi surface topology. Furthermore, we find that the superconducting transition temperature
is unaffected by this transition, demonstrating that their origins are different on the overdoped side.

\end{abstract}


\maketitle


Revealed more than twenty five years ago by nuclear magnetic resonance\cite{Alloul,Warren,Imai}, the pseudogap phase
in cuprates remains hitherto a mysterious state of matter out of which the high-temperature superconductivity emerges.
The pseudogap appears below the $T^{\ast}$ temperature and manifests itself as a loss of quasiparticle spectral
weight. Although intensely studied in the underdoped regime \cite{Timusk,Norman,Taillefer}, relatively less is known
about the pseudogap on the overdoped side, where it weakens and eventually disappears. Thus, a
logical line of enquiry is to study the pseudogap closing as a function of doping $p$, and to identify what
triggers it in the first place.

In systems where $T^{\ast}(p)$ intersects the superconducting dome described by the critical temperature
$T_c(p)$, this task is complicated by the appearance of the superconducting phase. One way to proceed
is to perform such a study at the lowest available temperatures,
either in the superconducting phase~\cite{Loram,Shekhter,Mangin-Thro,Fujita,He,Liu}, or by suppressing it with
magnetic field~\cite{Daou} or disorder\cite{Alloul2}. Often such studies have inferred a quantum phase
transition \cite{Varma,Tallon,Sachdev} associated with the pseudogap
closing.

A second possibility is to track the \emph{normal state} pseudogap at a higher temperature, and to study
the vicinity of the doping $p_c$ where it closes. Since $p_c$, defined as the doping where $T^{\ast}(p) = T_c(p)$,
is essentially a finite temperature property, \emph{a priori} it is not clear if it is linked to a quantum phase
transition.

In this work our main result is to show that in Bi$_{2}$Sr$_{2}$CaCu$_{2}$O$_{8+\delta}$ (Bi-2212) $p_c$
is indeed tied to a Lifshitz quantum phase transition where the underlying hole-like active Fermi surface
becomes electron-like at a van Hove singularity.
Interestingly, we find that $T_c$ is unaffected by this transition.
Moreover, comparing our results with existing photoemission and
tunnelling data of several hole-doped cuprates, we infer that
the microscopic origins of the pseudogap and the superconductivity are generically
different on the overdoped side. Only the former is tied to the change in the Fermi surface topology,
which removes quasiparticles from regions in momentum space of high scattering rate (hot regions).
While the collapse of the normal state pseudogap~\cite{Sacuto,Chatterjee,Vishik},
as well as the change of Fermi surface topology~\cite{Kaminski}
have been reported earlier, to the best of our knowledge,
the link between the two has not been demonstrated before in Bi-2212. Consequently, our result provides important clue
regarding the microscopic origin of the normal state pseudogap, and its relation with superconductivity.

One technical obstacle to study pseudogap closing is the lack of sufficiently overdoped samples belonging to the
same family of cuprates. Indeed, our study was made possible due to the availability of several
high quality Bi-2212 single crystals with doping close to $p_{c}$, as reported in Supplemental Material (SM).
The level of doping was controlled only by oxygen insertion, and the highest doping achieved was around
$p=0.24$.
This allowed us to perform a careful finely tuned electronic Raman study of the doping dependence
and determine $p_c = 0.22$.

The Raman measurements were performed in $\nu = B_{1g}, B_{2g}$ geometries
that probe respectively the antinodal (AN) region near $(\pm\pi,0)$ and $(0,\pm\pi)$ and the
nodal (N) region near $(\pm\pi/2,\pm\pi/2)$, (cf. SM).
Our spectra are comparable with earlier studies with a different laser line~\cite{Venturini1,Venturini2}, thereby
demonstrating absence of resonance effects in the overall conclusions.
In the following the quantity of importance is the integrated Raman intensity defined by
\begin{equation}
\label{eq:1}
I_{\nu}(T) = \int_0^{\Lambda} d \omega \chi^{\prime \prime}_{\nu} (\omega,T),
\end{equation}
extracted from the Raman response $\chi^{\prime \prime}_{\nu} (\omega,T)$ where $\Lambda$ is a cutoff.
We experimentally demonstrate our main result in two steps.

\begin{figure}[bp]
\begin{center}
\includegraphics[width=8.5cm,height=10cm]{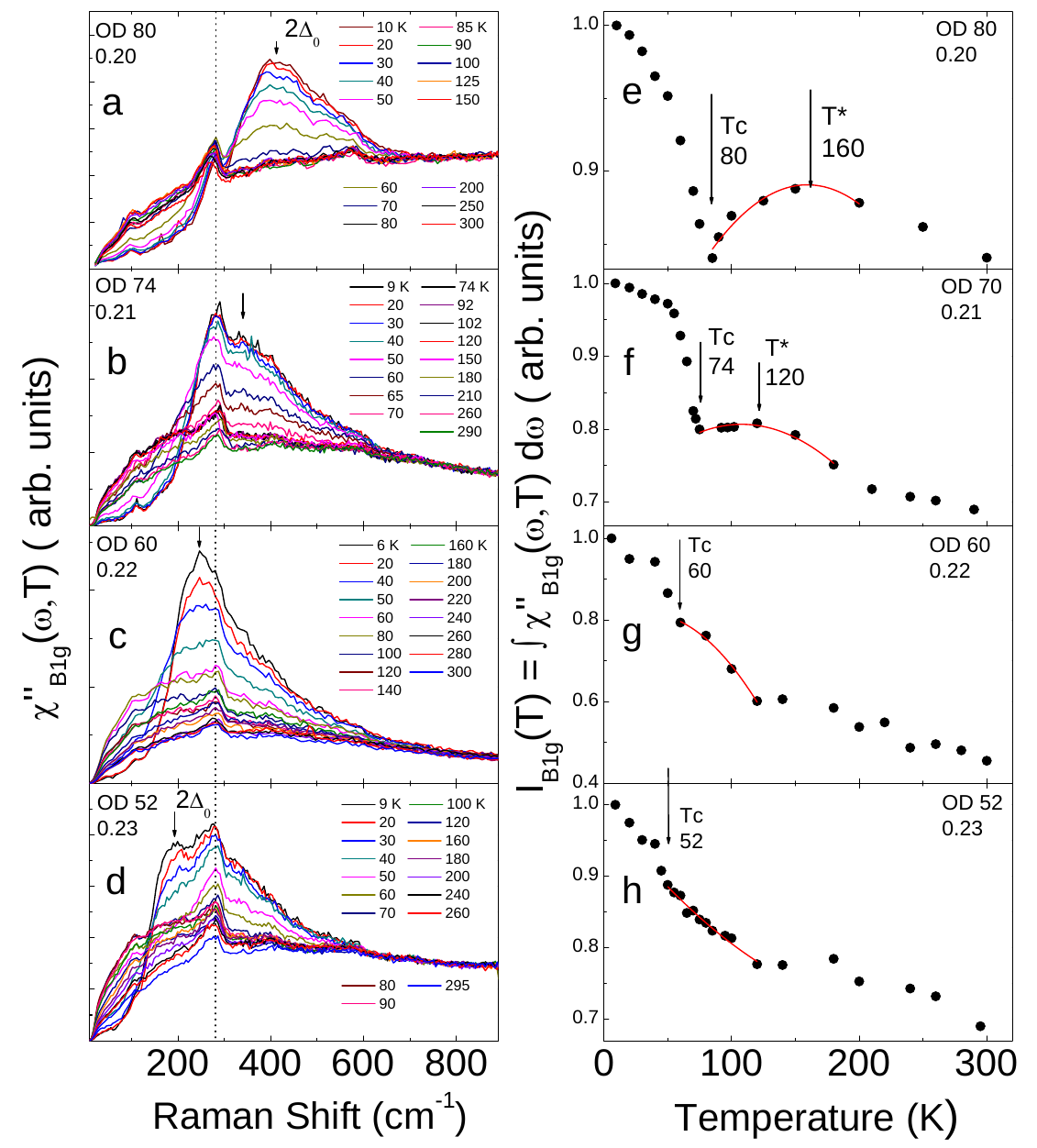}
\end{center}\vspace{-7mm}
\caption{(Color online)(a)-(d): $T$-dependence of the Raman response $\chi^{\prime \prime}_{\BAN}$
of overdoped Bi-2212. 2$\Delta_0$ is defined as the position of the \BAN pair breaking peak.
The location of the $300\,cm^{-1}$ phonon peak is marked by a dotted line.
(e)-(h): Integrated Raman intensities are
shown in (a)-(d), with cut off $\Lambda=850\,cm^{-1}$.
For each doping, $I_{\BAN}(T)$ is normalized by $I_{\BAN}(10 {\rm K})$.
The red curve is a second order polynomial fit just above $T_{c}$
indicating the sign change of the slope.
We found a remarkably linear dependence between $T_{c}$ and 2$\Delta_0$ in the critical
temperature range $90-50\,K$ (cf. SM). The doping level for each value of $T_{c}$ was fixed using the Tallon-Presland
formula (cf.SM).}
\label{fig1}
\end{figure}

In the first step we determine $p_c$ precisely,
which is a refinement of our earlier work~\cite{Sacuto}.
In Fig.~\ref{fig1}, (a)-(d), we report Raman responses $\chi^{\prime \prime}_{\BAN} (\omega, T)$ at
different temperatures and for different overdoped (OD) compounds.
The temperature dependence of the corresponding integrated intensities $I_{\BAN}(T)$ are shown in Fig.~\ref{fig1},(e)-(h).
In Fig. \ref{fig1}.a, we show the spectra for the Bi-2212 OD80 compound.
We observe that the pair-breaking peak ($2\Delta_0$), located at 408~cm$^{-1}$,
decreases in intensity with increasing temperature and disappears at $T_c$.
Correspondingly, $I_{\BAN}(T)$ decreases monotonically exhibiting a dip at $T_c$ (Fig.\ref{fig1}.e).
{Just above $T_c$ however, the low energy spectral weight (below 408~cm$^{-1}$) increases
(positive slope) with temperature.

This recovery of spectral weight,which can be as large as 15\% for $p=0.11$ (cf. Fig.~3 in SM),
is the signature of the presence of the pseudogap in the normal state spectra.
Note that, this $T$-dependence is opposite to that of a normal metal.
Therefore, above $T_c$, $I_{\BAN}(T)$ increases monotonically, until it reaches a maximum
at a temperature $T^{\ast}$ that defines the onset temperature of the pseudogap (Fig.\ref{fig1}.e).
Above $T^{\ast}$, the T-dependence of $I_{\BAN}$ is the one of a normal metal.
Our estimate of $T^{\ast}(p)$ is in good agreement with previous transport and spectroscopy
measurements (cf. Fig.~4 of SM).
As the doping level increases (Fig. \ref{fig1}. a-d), the difference between
$T_c$ and $T^{\ast}$ shrinks, and disappears at $p_c=0.22$, indicating the collapse of the normal state pseudogap.
For $p> 0.22$ the slope of $I_{\BAN}(T)$ just above $T_{c}$ is negative, implying there is no signature
of the normal state pseudogap anymore.
Quantitatively, this behavior is captured
by the doping dependence of the loss of the spectral intensity, defined by
$I_{\BAN}(T_{c})-I_{\BAN}(T^{\ast})$ and which we report in Fig.~\ref{fig3} as black stars.
Note that a change in the slope of $I_{\BAN}(T)$ at $p \geq 0.22$ appears at $T \approx 100$K
which is definitely higher than the pseudogap $T^{\ast} \approx 60$K. Consequently, this feature is
not related to the pseudogap, and its origin is currently under investigation.

\begin{figure}[htp]
\begin{center}
\includegraphics[width=8.5cm,height=10cm]{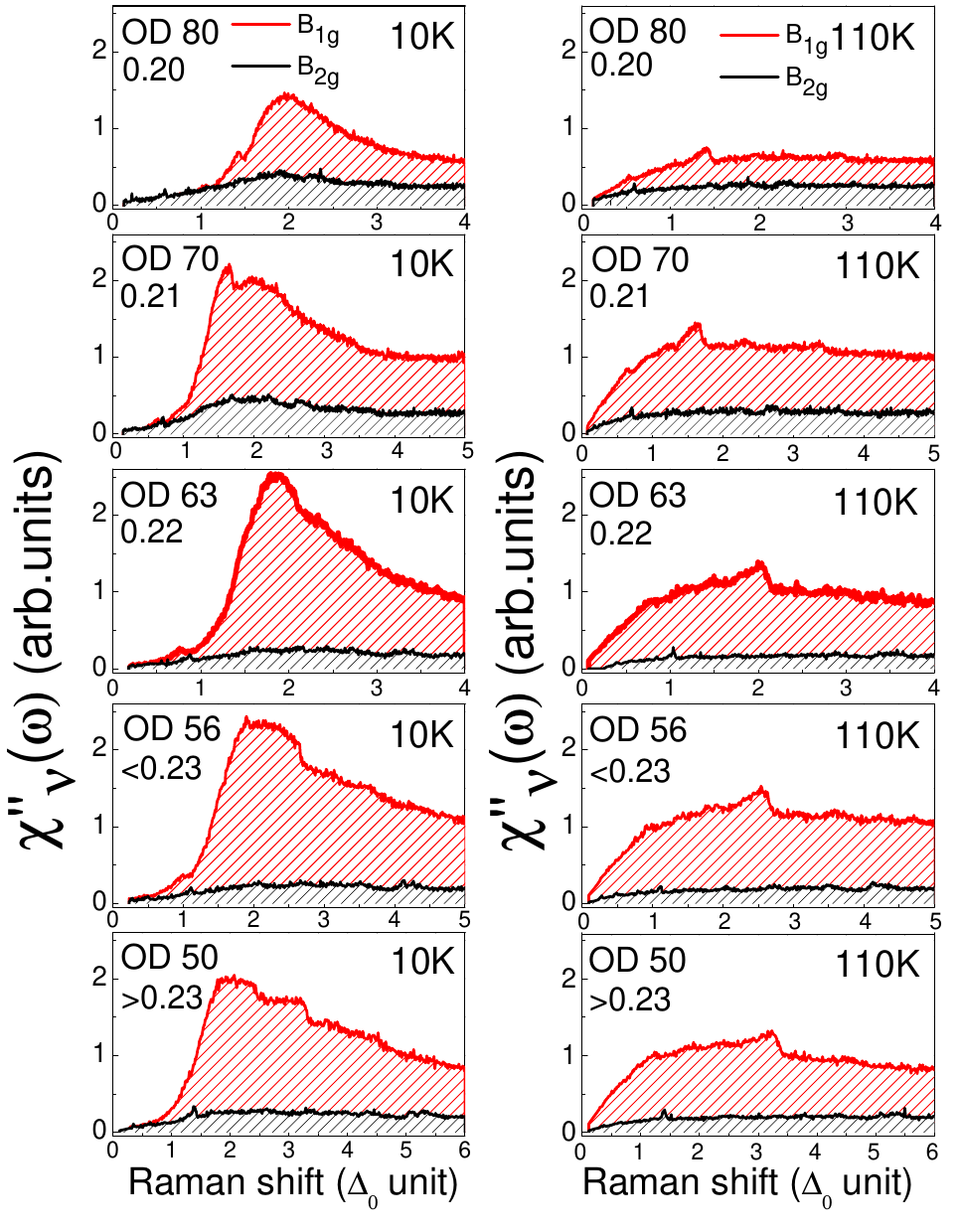}
\end{center}
\vspace{-7mm}
\caption{(Color online) \BAN (red/grey) and \BN (black) Raman responses of Bi-2212 at $10~K$ (superconducting state)
and $110~K$ (normal state) in the overdoped range using 532 nm laser.
The (red/grey) and black hatched areas indicate the magnitudes of the \BAN and the \BN responses respectively.
The former increases compared to the latter as a function of doping up to $p_c=0.22$.
}

\label{fig2}
\end{figure}

The second step involves comparing the spectra in the \BAN and \BN geometries, and following their doping evolutions
around $p_c$ at fixed temperatures. Importantly, we succeeded in measuring the \BAN and \BN Raman responses of each crystal on the same laser spot (see SM). We first show in Fig.~\ref{fig2} representative
$\chi^{\prime \prime}_{\BAN} (\omega)$ (red/grey) and $\chi^{\prime \prime}_{\BN} (\omega)$
(black) Raman responses, at $10~K$ (superconducting state) and at $110~K$ (normal state) for selected doping levels.
The Raman shift is expressed in units of the superconducting gap $\Delta_0 (p)$, in order to compare better samples
with varying gap values. Note  that, both in the superconducting and the normal states, the \BAN response increases
continuously in magnitude compared to the \BN response for $p \leq 0.22$, consistently with earlier
studies~\cite{Venturini1,Blanc,Hewitt,Munnikes,Masui1,Masui2,Naeini,Gasparov}.
However, the crucial finding of the current work is that this trend changes and  the \BAN response starts to decrease
beyond 0.22 doping.

\begin{figure}[htp!]
\begin{center}
\includegraphics[width=7cm,height=8cm]{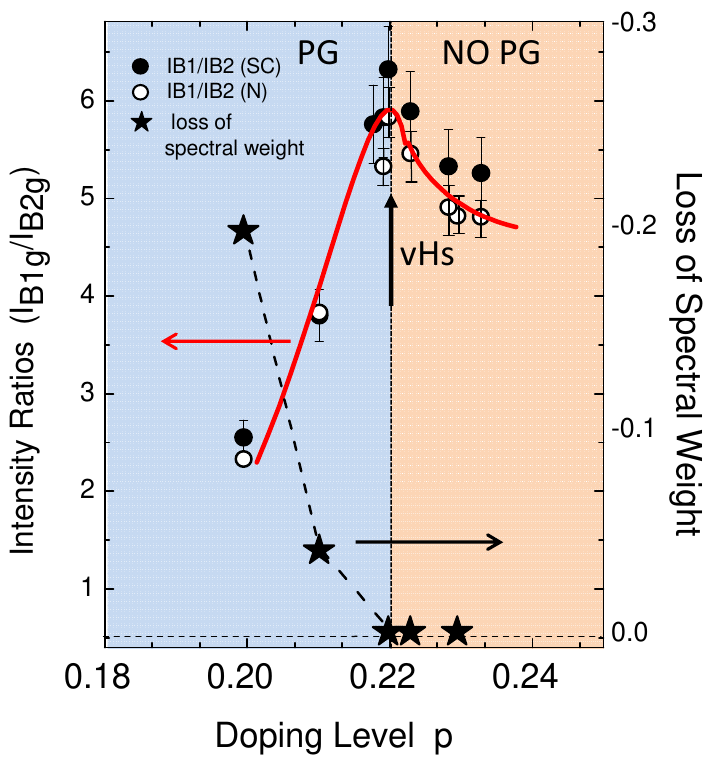}
\caption{(Color online)
Doping evolution of (i) the ratio $I_{B_{1g}}/I_{B_{2g}}$ of the integrated intensity (defined in
Eq.~\ref{eq:1}) in the superconducting and the normal states (filled and open circles respectively)
with cutoff $\Lambda \approx 3 \Delta_0$,
(ii) the loss of spectral weight related to the pseudogap (black stars).
The peak in the ratio $I_{B_{1g}}/I_{B_{2g}}$, both for the superconducting and the normal phases,
coincides with the critical doping $p_c=0.22$ where the pseudogap disappears. The peak is a consequence of a Lifshitz transition where the hole-like Fermi surface
of the dominant anti-bonding band becomes electron-like as the chemical potential crosses a van Hove singularity.}
\label{fig3}
\end{center}\vspace{-7mm}
\end{figure}


The above non-monotonic doping dependence is best quantified by extracting the ratio of the integrated
intensities $I_{B_{1g}}/I_{B_{2g}}$ from the Raman responses $\chi^{\prime \prime}_{\nu} (\omega)$
using Eq.~\ref{eq:1}.
Studying the intensity ratio, rather than the absolute intensities,
allow us to avoid spurious effects due to non intrinsic intensity modulations that may occur when passing
from one crystal to another (cf.SM). Note that, in principle $I_{\BAN}$ contains not only the
contribution of the electronic background (which is what we are interested in), but also that of the phonon
peaked sharply at 300 cm$^{-1}$ (see the dotted line in Fig.~\ref{fig1}). However,
by comparing the current spectra with that obtained using 647.1 nm laser line, in which the phonon peak is absent,
we are able to confirm that $I_{\BAN}$, and especially its doping dependence, is mostly due to the electronic
background \cite{note1}.

In Fig.~\ref{fig3} we report the doping dependencies of the intensity ratios $I_{B_{1g}}/I_{B_{2g}}$
in the superconducting (filled circles) and the normal states (open circles). Note that, the ratios
in the two phases are nearly the same, thereby indicating that $I_{B_{1g}}/I_{B_{2g}}$ is unaffected
by the superconducting gap.
Most importantly, the ratios change non-monotonically as a function of $p$, and they reveal
a sharp peak located at $p_c =0.22$, the doping where the normal state pseudogap closes (black stars).
We confirmed that the sharp peak is not a resonance effect, since it is visible with two
distinct laser lines (532 nm and 647.1 nm).

Note that the peak in $I_{B_{1g}}/I_{B_{2g}}$ cannot be attributed to the doping dependence of the pseudogap which is
monotonic. Instead, the temperature independence of the sharp peak position indicates that it is related to enhanced
density of states of the underlying band structure around the AN region of the Brillouin zone.
This invariably leads to the possibility of a doping induced Lifshitz transition wherein,
as a van Hove singularity crosses the chemical potential, the open hole-like anti-bonding Fermi surface
closes around the $(\pm \pi,0)$ and $(0, \pm \pi)$ points and becomes electron-like.
An electron-like anti-bonding band in Bi-2212 at $p>0.22$ has been reported by ARPES
data~\cite{Kaminski}, but this change of topology was not linked with the closing of the pseudogap.

\begin{figure}[tp]
\begin{center}
\includegraphics[width=8.5cm,height=7cm]{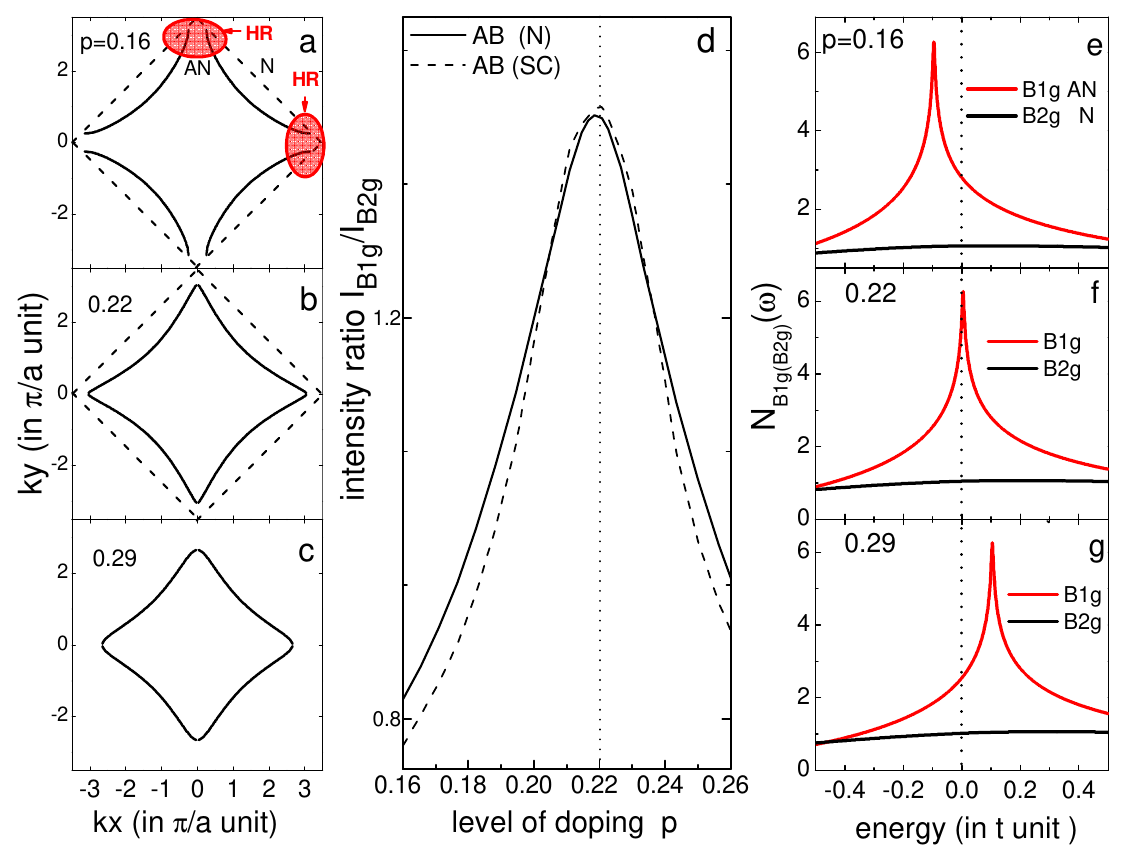}
\end{center}\vspace{-7mm}
\caption{(Color online)
(a)-(c) With doping hole-like anti-bonding Fermi surface becomes electron-like at a Lifshitz transition.
The dotted line is the antiferromagnetic Brillouin zone. HR denotes the hot region of large scattering rate.
(d) Doping dependence of the integrated intensity ratio $I_{B_{1g}}/I_{B_{2g}}$ in the normal (solid line)
and superconducting (dashed line) states calculated from a theoretical model (see text), reproducing
qualitatively the trend of Fig.~\ref{fig3}.
The curves are normalized (cf.SM). (e)-(g). The associated van Hove singularity appears in $N_{B_{1g}}(\omega)$ (red/grey),
the density of states weighted by the \BAN Raman vertex (defined in text), but not in $N_{B_{2g}}(\omega)$ (black),
which is multiplied by $(t/t')^2$ for better visibility.
}
\label{fig4}
\end{figure}

In order to support this scenario we perform theoretical calculation of the Raman response function using a minimal
tight-binding model

with the normal state dispersion~\cite{Kordyuk}:
$
\epsilon_{k,\alpha}=-2t(\cos k_x + \cos k_y)+ 4t'\cos k_x \, \cos k_y \pm t_o \, (\cos k_x - \cos k_y)^2 /4-\mu.
$
Here $\alpha=\pm$ refer to the anti-bonding (AB) and the bonding (B) bands.
The superconducting dispersion is

$E_k = \sqrt{\epsilon_k^2 + \Delta_k^2}$, with $ \Delta_k= \Delta_0( \cos k_x- \cos k_y )/2$.
We take $t'= -0.3 t$, $t_o= 0.084 t$, and a doping independent $\Delta_0=0.0025t$. We change $p$ by
varying the chemical potential $\mu$.
As shown in Fig.~\ref{fig4}(a)-(c), this model undergoes a Lifshitz transition at $p_c=0.22$ where the AB band changes
from being hole- to electron-like (the B band remains hole-like in this doping range, see SM).
For simplicity we take a constant electron scattering rate $\Gamma_N =0.01t$
and $\Gamma_S = 0.0025t$ in the normal and the superconducting states respectively.
An earlier work has shown that the scattering rates measured from the slopes of the Raman responses become
isotropic around  $p \approx 0.22$~\cite{Venturini1}.
The calculation of $\chi^{\prime \prime}_{\nu}(\omega)$ and $I_{\nu}$ are standard
(for details, cf. SM). The doping dependence of the calculated ratio $I_{B_{1g}}/I_{B_{2g}}$ shows
prominent peaks at $p=0.22$ (see Fig.~\ref{fig4} (d)), both in the normal and the superconducting states,
and reproduces qualitatively the experimental trend of Fig.~\ref{fig3}.

The origin of the peak can be captured conveniently by tracking the doping dependence of the Raman vertex
$\gamma_{k,\alpha}^{\nu}$-weighted density of states
$N_{\nu}(\omega) \equiv \sum_{k,\alpha} (\gamma_{k,\alpha}^{\nu})^2 \delta(\omega - \epsilon_{k,\alpha})$
which enter the calculation of $I_{\nu}$. As shown in Fig.~\ref{fig4}(e)-(g), the van Hove singularity shows up in
$N_{B_{1g}}(\omega)$, and the peak in the intensity $I_{B_{1g}}$ corresponds to the van Hove singularity crossing the
chemical potential. Simultaneously, since the
$B_{2g}$ geometry probes the diagonal directions of the Brillouin zone, $N_{B_{2g}}(\omega)$ is unaffected by the
van Hove singularity and therefore $I_{B_{2g}}$ has no significant doping dependence.

Based on the Mott formula, at the Lifshitz transition $p=0.22$
we expect a change in the sign of the Seebeck coefficient,
provided the scattering rates are isotropic~\cite{Buhmann2013}. However, as noted in an earlier work \cite{Kaminski},
the Hall coefficient may remain positive across the Lifshitz transition,
as observed in thin film studies~\cite{Konstantinovic}.

A possible interpretation of our results is that the Lifshitz transition avoids the pseudogap by
effectively moving quasiparticles from regions of strong scattering (hot regions) located around
$(\pm \pi,0)$ and $(0, \pm \pi)$ (see panels (a)-(c), Fig.~\ref{fig4}). Note that, this possibility
is independent of the origin of the hot regions which could arise
from fluctuations of antiferromagnetic spin-waves ~\cite{Chubukov1,Schmalian,Ioffe,Chakravarty}
or charge-density waves (related to long-ranged incommensurate charge
modulations)\cite{Chubukov2,Comin,Neto,Julien,Sebastian,Pepin},
or from Mott-related physics~\cite{Civelli,Kyung,Ferrero,Sakai,Sordi}.
A second quantum critical point at $p*$ inside the superconducting dome, separating a small from a large FS phase, has been
inferred for example in recent scanning tunnelling microscopy~\cite{Fujita,He} and ARPES~\cite{Vishik}
experiments.

\begin{figure}[t]
\begin{center}
\includegraphics[width=9cm]{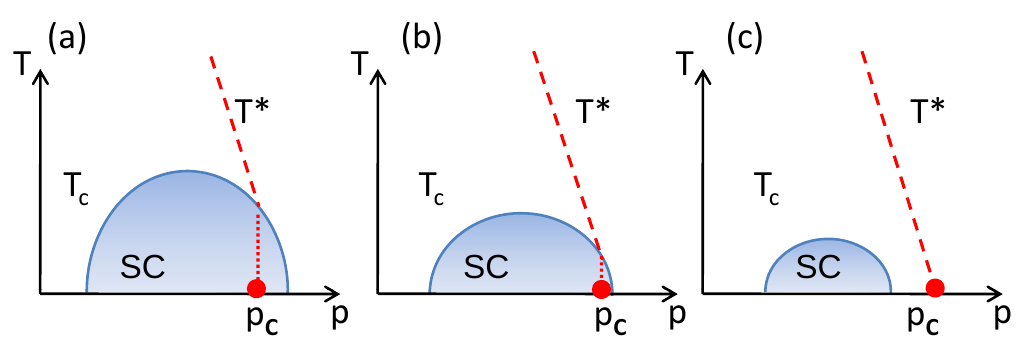}
\caption{(Color online) Schematic temperature ($T$) versus doping $p$ phase diagram with three
material-dependent possible locations of the critical doping $p_c$, where the normal state pseudogap
closes, with respect to the superconducting (SC) dome (shaded blue/grey). (a) is realized in Bi-2212
(current study),
(b) and (c) have been reported for LSCO ~\cite{Ino} and Bi-2201 ~\cite{Piriou} respectively. The common feature in all three cases
is the coincidence of $p_c$ with a Lifshitz transition where a Fermi surface changes from hole-like to electron-like.}
\label{fig5}
\end{center}
\end{figure}

An intriguing pattern emerges upon comparing our results with those on other hole doped cuprate families near $p_c$. In contrast with Bi-2212 and La$_{1.6-x}$Nd$_{0.4}$Sr$_x$CuO$_4$ \cite{Daou}
where $p_c$ is located well inside the superconducting dome, in ARPES measurements
on La$_{2-x}$Sr$_x$CuO$_4$ (LSCO) of Ref. \cite{Ino} the endpoints of the pseudogap and the superconducting phases
are nearby in doping. Scanning tunneling spectroscopy on Bi$_{2}$Sr$_{2}$CuO$_{6+\delta}$,(Bi-2201)
instead, found the pseudogap extending well into the normal phase~\cite{Piriou}.
This suggests that the position of $p_c$ with respect to the superconducting dome
is material dependent (see Fig.~\ref{fig5}).
Interestingly, just as we established here for Bi-2212, for both LSCO and
Bi-2201 data analyses have suggested the coincidence of the pseudogap closing
with a Lifshitz transition~\cite{Ino,Piriou}.
Taken together, this appears to be a universal feature of the hole doped cuprates,
and our findings establish an intimate connection between the normal state pseudogap
and Fermi surface topology. In Tl$_{2}$Ba$_{2}$CuO$_{6+\delta}$ (Tl-2201) the scenario
is less clear, as the observation of the pseudogap is still debated~\cite{Damascelli}.

Few studies \cite{Ando,Tallon}
on Bi-2212 and YBa$_2$Cu$_3$O$_{7-x}$,(Y-123) have reported the pseudogap closing at $p=0.19$. This might simply
imply that the normal state and the superconducting pseudogaps close at different dopings~\cite{Vishik}.
Alternately, this apparent discrepancy could be related to the fact that in-plane transport and superfluid density
are mostly sensitive to the nodal properties~\cite{Ioffe}, while $c$-axis transport and \BAN Raman probe mostly the antinodal
properties~\cite{Andersen}. Next, in our scenario it is possible that for $p>0.22$ the pseudogap exists
in the hole-like B band, but we do not find any signature of it in the Raman spectra, consistently
with ARPES results~\cite{Vishik}. One possibility is that
the response is predominantly from the AB band since it is close to a density
of states singularity. We notice this trend in the theoretical calculation as well (cf. SM).

In conclusion, our results demonstrate that the mechanism that gives rise to the normal
state pseudogap is sensitive to the topology of the Fermi surface, and is operational only
when the latter is hole-like. Furthermore, we conclude that,
on the overdoped side of the cuprates, the microscopic origins
of the pseudogap and the superconductivity are different.

We are grateful to A. ~Georges, J. Tallon, A. Chubukov, M. Norman, L. Taillefer, C. P\'epin, Ph. Bourges,
Y. Sidis and A. Damascelli for very helpful discussions. Correspondences and requests for materials should be addressed to
A.S (alain.sacuto@univ-paris-diderot.fr)


\section*{SUPPLEMENTAL MATERIAL}

\subsection*{A. Raman Experiments}
Raman experiments have been carried out using a triple grating spectrometer (JY-T64000) equipped with a
liquid-nitrogen-cooled CCD detector.
Two laser excitation lines were used: 532 nm and 647.1 nm from respectively a diode pump solid state laser and
a Ar+/Kr+ mixed laser gas.
The \BAN and \BN  geometries were
obtained from cross polarizations at 45$^o$ from the Cu-O bond directions
and along them
respectively. The change from \BAN to \BN geometries was obtained by keeping fixed the orientations of the analyzers
and the polarizers
and by rotating the crystal using an Attocube piezo-driven rotator. We  got an accuracy on the crystallographic
axes orientation with
respect to the polarizers close  to $2^o$. Importantly, we succeeded in measuring the \BAN and \BN Raman responses
of each crystal on the same laser spot. This allowed us to keep constant the solid angle of collection and made
reliable the \BAN to \BN Raman integrated intensity ratio. Studying the intensity ratio rather than the absolute
intensities of the Raman response prevent us from some non intrinsic intensity modulations when passing from one crystal
to another.

All the spectra have been corrected for the Bose factor and the instrumental spectral response. They are
thus proportional to the imaginary part of the Raman response function $\chi^{\prime \prime}_{\nu} (\omega)$.
Measurements between $4$ K and $300$ K have been performed using an ARS closed-cycle He cryostat. The laser power
at the entrance of the cryostat was maintained below $2~mW$ to avoid over heating of the crystal estimated to
$3$ K/mW at $10$ K.

\subsection*{B. Crystal Growth and Characterization}

The Bi-2212 single crystals were grown by using a floating zone method. The optimal doped sample with $T_{c} = 90~K$
was grown at a velocity of 0.2 mm per hour in air ~\cite{Wen_b}. In order to get overdoped samples down to $T_{c}=65~K$ ,
the as-grown single crystal was put into a high oxygen pressured cell between $1000$ and $2000$ bars and then was
annealed from $350^{o}C$ to $500^{o}C$ during 3 days ~\cite{Mihaly}.
The overdoped samples below $T_{c}=60~K$ was obtained from as-grown Bi-2212 single crystals put into a pressure
cell (Autoclave France) with $100$ bars oxygen
pressure and annealed from $9$ to $12$ days at $350~^{o}C$. Then the samples were
rapidly cooled down to room temperature by maintaining a pressure of $100$ bars. The critical temperature $T_{c}$
for each crystal has been
determined from magnetization susceptibility measurements at a $10$ Gauss field parallel to the c-axis of the crystal.
More than 30 crystals
have been measured among $60$ tested. The selected crystals exhibit a quality factor of $T_{c}/ \Delta T_{c}$ larger
than $7$.
$\Delta T_{c}$ is the full width of the superconducting transition. A complementary estimate of $T_{c}$  was
achieved from electronic Raman scattering measurements by defining the temperature from which the \BAN superconducting
pair breaking peak collapses.

\subsection*{C. Estimate of $\Delta_{0}$ and its relationship to $T_{c}$ }

The determination of $\Delta_{0}$ was achieved by subtracting the normal \BAN  Raman response at $110~K$ from the
\BAN one in the superconducting state at $10~K$. We define $2\Delta_{0}$  as the maximum of the electronic background
in the subtracted spectra, see Fig.1.

\begin{figure}[ht!]
\begin{center}
\includegraphics[width=1\linewidth]{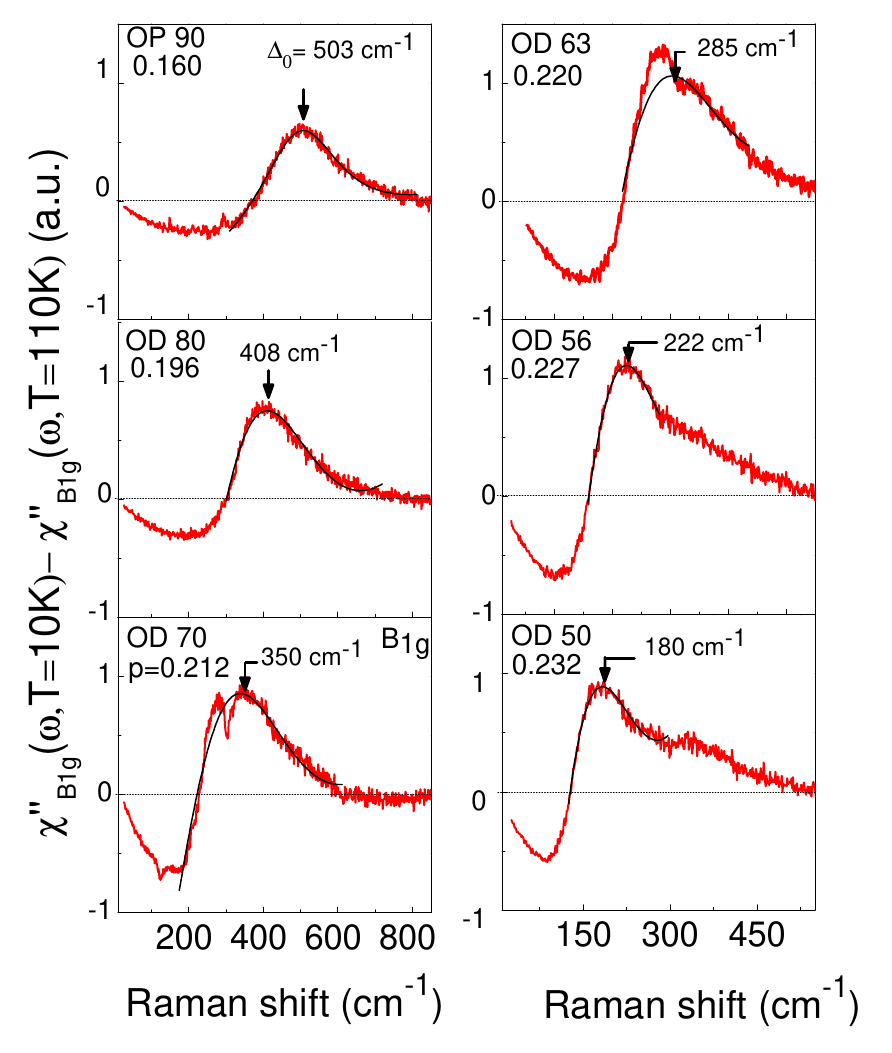}
\end{center}\vspace{-7mm}
\caption{Subtraction between the normal and superconducting \BAN Raman responses of Bi-2212 for several doping levels.
The arrows indicate the location of $2\Delta_{0}$ in wavenumbers. An independent accurate estimate of
$2\Delta_{0}$ for OD 70 and
OD 63 compounds was obtained using the 647.1 nm laser line (not shown).
For this excitation line the phonon peak
at $283~cm^{-1}$ is no more Raman active and it does not hamper anymore the estimate of $2\Delta_{0}$.}
\label{fig6}
\end{figure}

Special care has been devoted to select single crystals which exhibit the same $2\Delta_0$ value in the Raman spectra
measured from distinct laser spots on a freshly cleaved surface.
We find that $T_{c}$ increases linearly with $2\Delta_0$ in the overdoped regime and reaches its maximum value
at $T_{c}^{max}=90~K$. From a linear fit of the $T_c$ values in a short range between $T_c=50\,K$ and $T_c=90\,K$,
we find the reliable relationship:  $T_{c}=(2\Delta_{0})/8.2+28.6$. In the underdoped regime $T_{c}$ falls down
abruptly as a function of $2\Delta_0$ (see Fig.~\ref{fig7}). The level of doping $p$ was defined from $T_c$
using Presland and Tallon's equation\cite{Presland}: $1-T_{c}/T_{c}^{max} = 82.6 (p-0.16)^{2}$. In the overdoped regime,
estimate of $p$ can be directly determined from $2\Delta_0$ using the above two equations.

\begin{figure}[ht!]
\begin{center}
\includegraphics[width=1\linewidth]{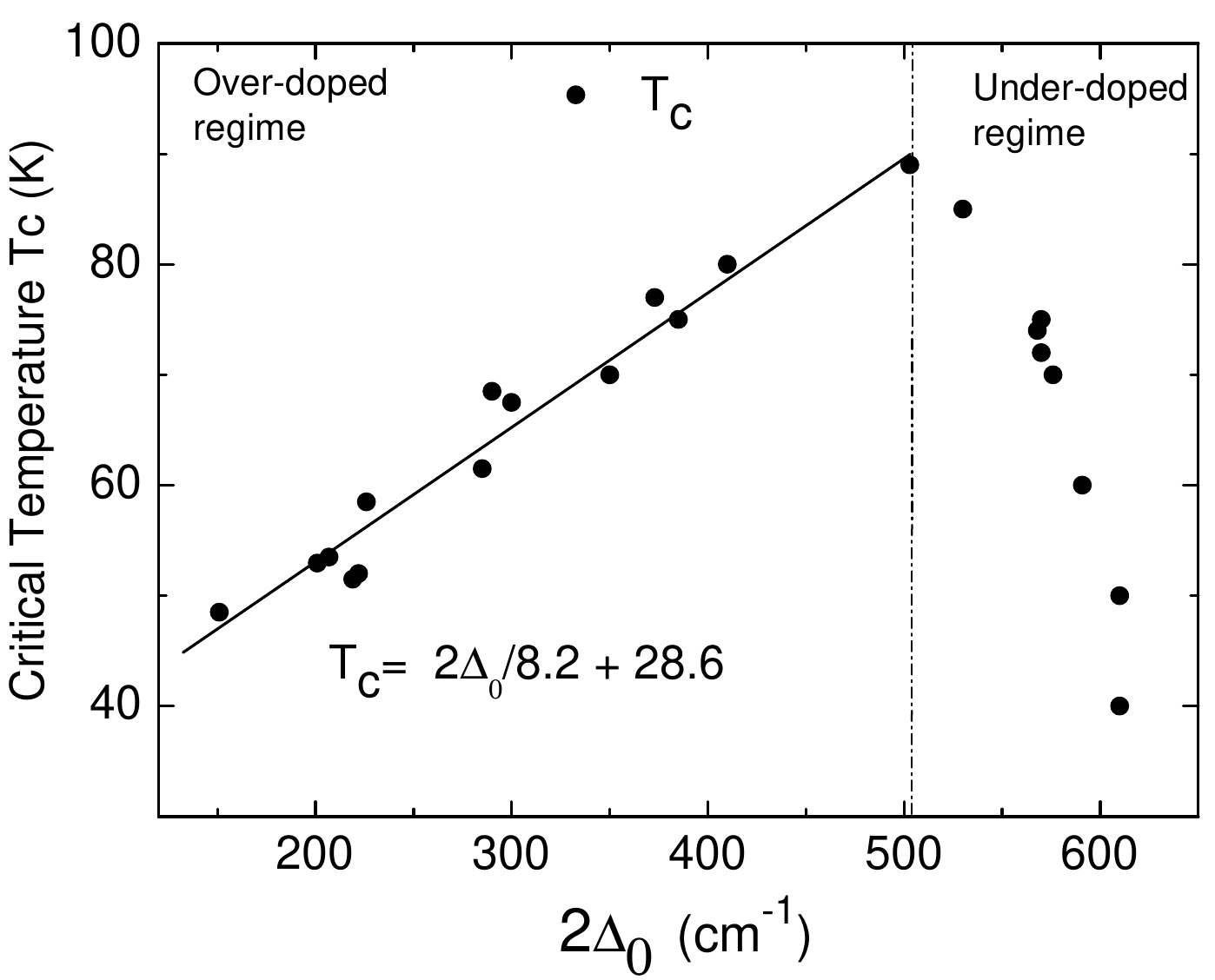}
\end{center}\vspace{-7mm}
\caption{Evolution of the critical temperature $T_c$ versus pair breaking peak $2\Delta_{0}$.
$T_c$ and $2\Delta_{0}$ are respectively deduced from magnetic susceptibility and  Raman measurements of Bi-2212 crystals with distinct oxygen annealing treatments (see text). The black solid line corresponds to the linear fit of $T_c$ in a restricted range of the overdoped side. The dashed-doted line marks off the end of the overdoped regime. The maximum value of $T_c$ is reached for $90~K$ which corresponds to $2\Delta_{0}=~503~cm^{-1}$.}
\label{fig7}
\end{figure}
\subsection*{D. Signature of the pseudogap from \BAN integrated Raman Intensity}

In Fig.\ref{fig10} is reported the temperature dependence of the \BAN Raman response and its integrated Raman
intensity for several doping levels.
\begin{figure}[!h]
\begin{center}
\includegraphics[width=1\linewidth]{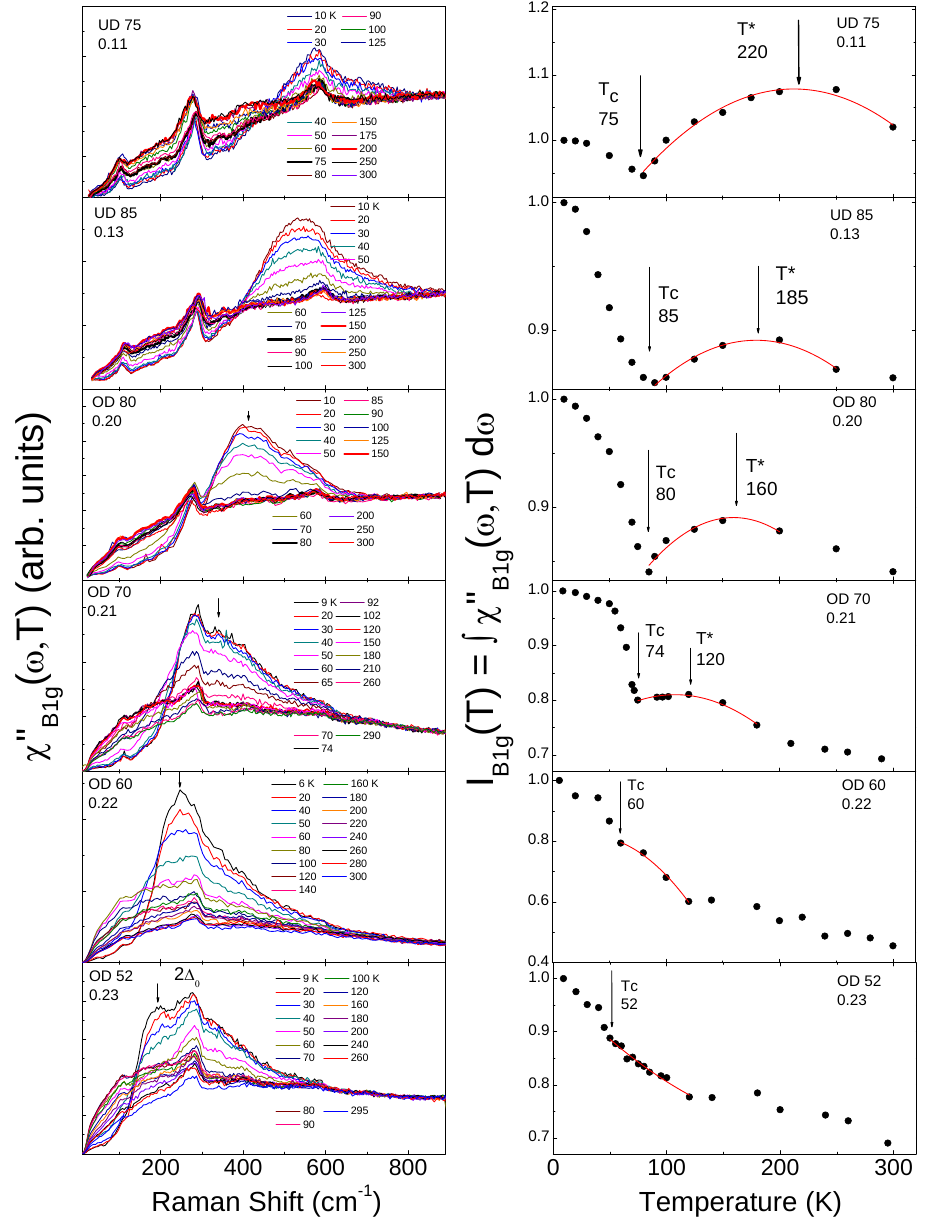}
\end{center}\vspace{-7mm}
\caption{Temperature dependence of the Raman response $\chi^{\prime \prime}_{\BAN}$ and its  integrated Raman
intensity  $I_{\BAN}(T)$  for underdoped and overdoped Bi 2212 single crystals. For each doping level,
$I_{\BAN}(T)$ has been normalized to this value at $T=10\,K$. Note that for p=0.11 (UD 75 crystal)
the pseudogap signal is large and represents $15\%$ of the integrated Raman intensity at $T=10K$. Consequently
its doping dependence can be easily tracked.}
\label{fig10}
\end{figure}
Below $p=0.22$ the Raman integrated intensity curves exhibit a dip at $T_{c}$ and reaches a maximum at
$T_{\ast}$ that defines the onset of the pseudogap. The loss of spectral intensity
$I_{\BAN}(T_{c}) - I_{\BAN}(T^{\ast})$
is taken as the strength of the pseudogap. For p=0.11 (UD 75 crystal, Fig.~\ref{fig10}, first row)
the strength of the pseudogap is clearly detected
and represents $15\%$ of the integrated Raman intensity at $T=10K$.
Importantly, the signature of the pseudogap manifests itself by a positive slope of $I_{\BAN}(T)$ just
above $T_{c}$.This slope stays positive for $p \leq 0.22$, and changes sign only above $p_c = 0.22$.
This change of sign has been underlined by fitting $I_{\BAN}(T)$ with a second order
polynomial function (red curve in Fig.\ref{fig10}) just above Tc. The absence of the pseudogap is then characterized
by a negative slope of $I_{\BAN}(T)$ just after $T_{c}$.

\subsection*{E. Comparison between $T^{\ast}$ determined from Raman and c-axis transport and other spectroscopy
probes}
It is important to point out that the doping evolution of $T^{\ast}$ from our Raman results on Bi-2212 is in good agreement
with angle resolved photoemission (ARPES), recent neutron scattering, tunneling and resistivity measurements along
the c-axis.
This is displayed in Fig.\ref{fig11} and show that the electronic Raman scattering is a reliable probe for detecting the
pseudogap and for studying its doping evolution.
Note that $T^{\ast}$ got from \BAN Raman response has been compared to the c-axis tunneling and transport measurements
because all these experiments probe mostly the anti-nodal properties (cf. O. K. Anderson, J. Phys. Chem. Sol. {\bf 56},
1573, 1995).


\begin{figure}[!h]
\begin{center}
\includegraphics[width=1\linewidth]{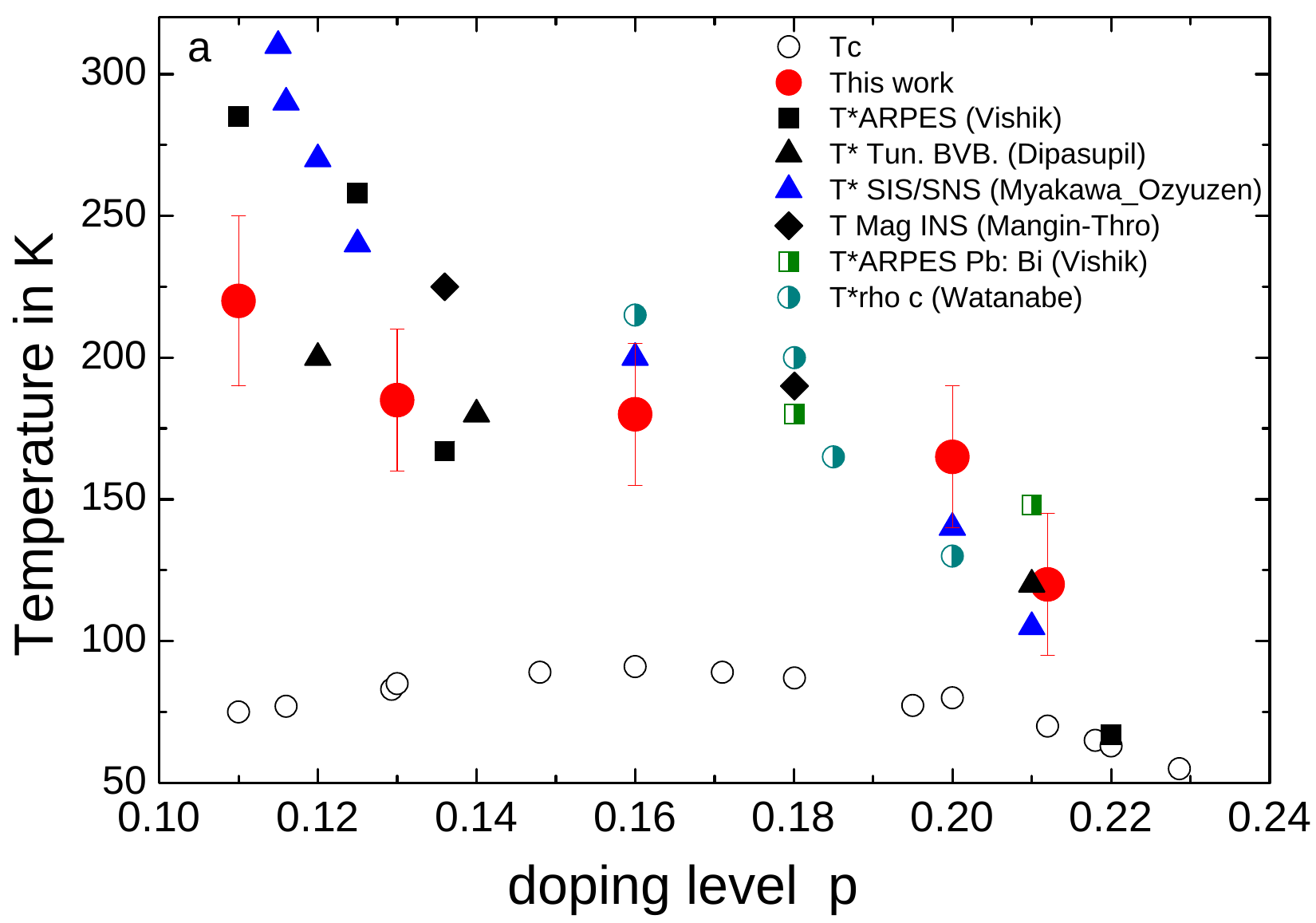}
\end{center}\vspace{-7mm}
\caption {Comparison between $T^{\ast}$ determined by Raman (filled red circles)
and transport and other spectroscopic probes. At $p_c=0.22$ there is no signature of normal state pseudogap
(see Fig.~\ref{fig10}), consequently $T^{\ast}(p_c) = T_c(p_c)$.
The references are (i) tunneling : R. M. Disapusil et al. J. Phys. Soc. Jpn. {\bf 71}, 1535, (2002), L. Ozyuzer et al., Eur. Phys. Lett.
{\bf 58}, 589, (2002), N. Miyakawa et al., Phys. Rev. Lett. {\bf 83}, 1018, (1999). (ii) resistivity : T. Watanabe et al., Phys. Rev. Lett. {\bf 84},
5848, (2000); T. Usui et al. arXiv:1404.473, (iii)ARPES: I.M. Vishik et al., Proc. Natl. Acad. Sci. {\bf 109}, 18332, (2012).
(iv) neutron scattering: L.Mangin-Thro et al., Phys. Rev. B  {\bf 89}, 094523, (2014).}
\label{fig11}
\end{figure}
\newpage

\subsection*{F. Theoretical Raman calculation}

We calculate the electronic Raman response using standard zero-temperature linear response theory \cite{Devereaux,Chubukov}.
This is given by

\begin{align}
\label{eq:chi}
\chi^{''}_{\nu}(\Omega) =& \sum_{\alpha= AB, B} \, \int^{0}_{-\Omega} \, \sum_{k} (\gamma^{\nu}_{k,\alpha})^2
\left[ \hbox{Im}G_{\alpha}(k,\omega)  \right. \nonumber \\
\times& \left. \hbox{Im} G_{\alpha}(k,\omega+\Omega)  - \hbox{Im}F_{\alpha}(k,\omega)\hbox{Im}
F_{\alpha}(k,\omega+\Omega) \right],
\end{align}
where the Raman vertex in the geometry $\nu = (B_{1g}, B_{2g})$ is
$\gamma^{\nu}_{k,\alpha}=\frac{m}{\hbar^2}\sum_{a,b}e^{I,\nu}_{a}\frac{\partial^2\epsilon_k^{\alpha}}
{\partial k_a \partial k_b}e^{S,\nu}_{b}$, with $(a,b)$ denoting spatial directions $(x,y)$,
$e^{I}$ and $e^{S}$ are polarization vectors for the incident and the scattered light respectively.
The normal-state one particle propagator is given by
$G(k,\omega)= 1/(\omega- \varepsilon_k+ \hbox{i} \Gamma_N)$, while in the superconducting state it can be written in a
Nambu formalism:
\begin{eqnarray}
\label{eq:G}
\hat{G}(k,\omega)=
\left(
\begin{array}{cc}
G(k,\omega)   &  F(k,\omega) \\
 F(k,\omega)      & -G^{*}(k,-\omega)  \end{array}
\right)
= \nonumber \\
\left(
\begin{array}{cc}
 \omega- \varepsilon_k+ \hbox{i} \Gamma_S  & \Delta_k  \\
\Delta_k       &   \omega+ \varepsilon_k+ \hbox{i} \Gamma_S \end{array}
\right)^{-1}
\end{eqnarray}

Here we have introduced constant scattering rates $\Gamma_N= 0.01 t$ and $\Gamma_S=0.0025 t$ in the normal and
superconducting states respectively.

The superconducting gap is given by
$\Delta_k= \, (\Delta_0/2) (\cos k_x - \cos k_y)$, with $\Delta_0= 0.025 t$. Since our aim
is to study only the effect of the Fermi surface topology change and the associated van Hove singularity
on the Raman response, we do not change $\Delta_0$ with doping.

The Lifshitz transition in the AB band at the critical doping $p_c=0.22$ is clearly visible in  the \BAN Raman
response,
since in this geometry one probes mainly the antinodal parts of the Fermi surface close to the $k= (0,\pm \pi)$
and $(\pm \pi, 0)$. The
Raman-vertex-weighted density of states $N_{\BAN}(\omega)$ (defined in the main text) presents in this case a
singularity at the chemical potential (see Fig. 3.(f), main text), and therefore the $\chi^{''}_{\BAN}(\omega)$
increases substantially and present a maximum at $p_c$, see top panels of Fig.~\ref{fig8} (a) and (b). Instead, the
$N_{\BN}(\omega)$ does not present any singularity, since in this geometry one probes mainly the nodal regions
of the Brillouin zone.
Consequently the \BN Raman response is not affected by the Lifshitz transition, and it displays little doping
dependence.

\begin{figure}[htp]
\begin{center}
\includegraphics[width=8cm,height=10cm]{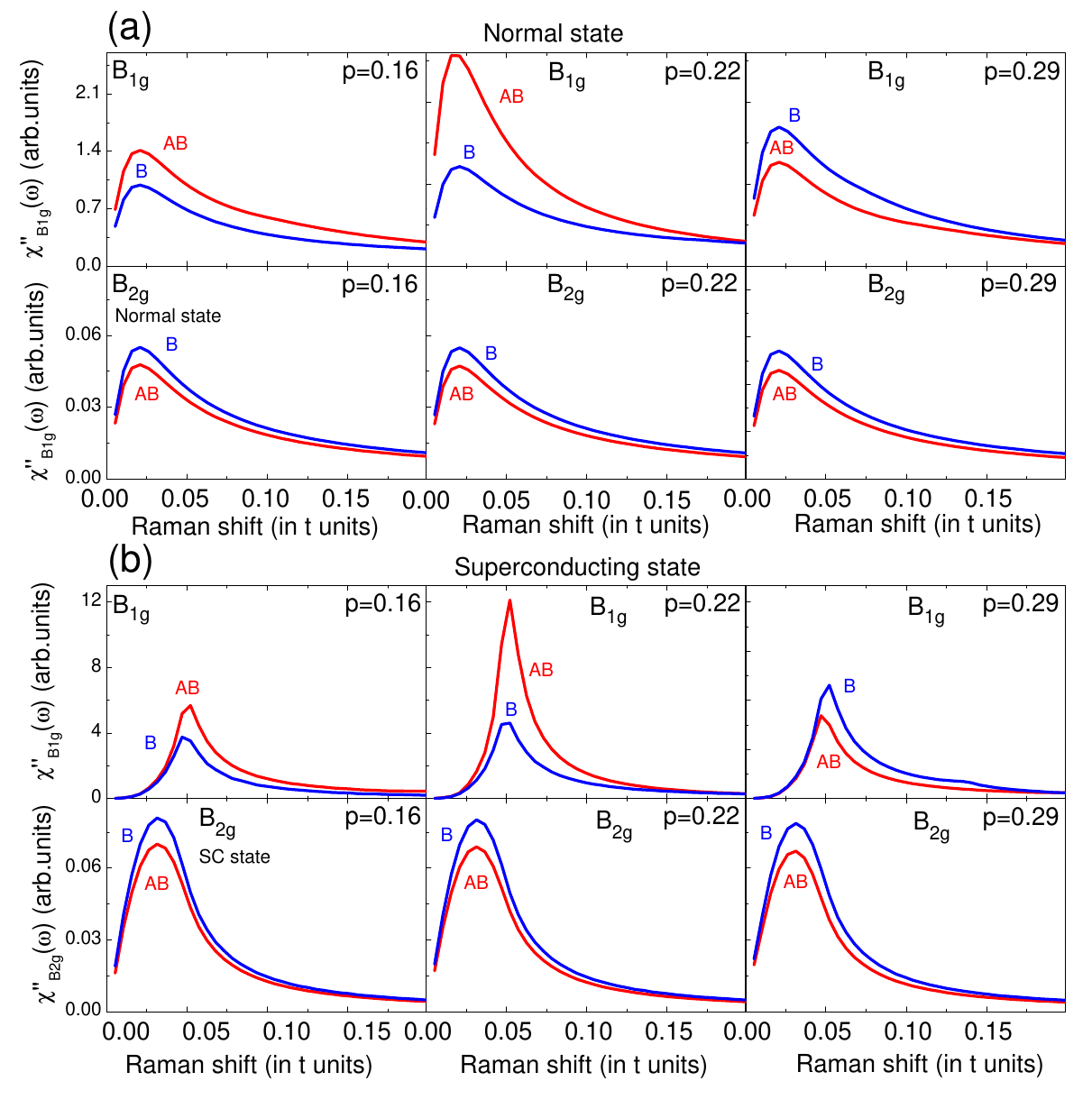}
\end{center}
\caption{Raman response evolution (a) in the normal state and (b) superconducting state  for three characteristic dopings.
The doping level $p=0.22$ corresponds to a Lifshitz transition where the hole-like
AB band becomes electron-like. The AB band contribution is represented by (red/light gray) lines and the B band
contribution by (blue/dark gray) lines.
Both in the normal and the superconducting states the \BAN response of the AB band shows a substantial enhancement at
the van Hove singularity associated with its Lifshitz transition.
The  B$_{2g}$ response, by contrast, is only weakly doping dependent.}
\label{fig8}
\end{figure}

\begin{figure}[htp]
\begin{center}
\includegraphics[width=0.8\linewidth]{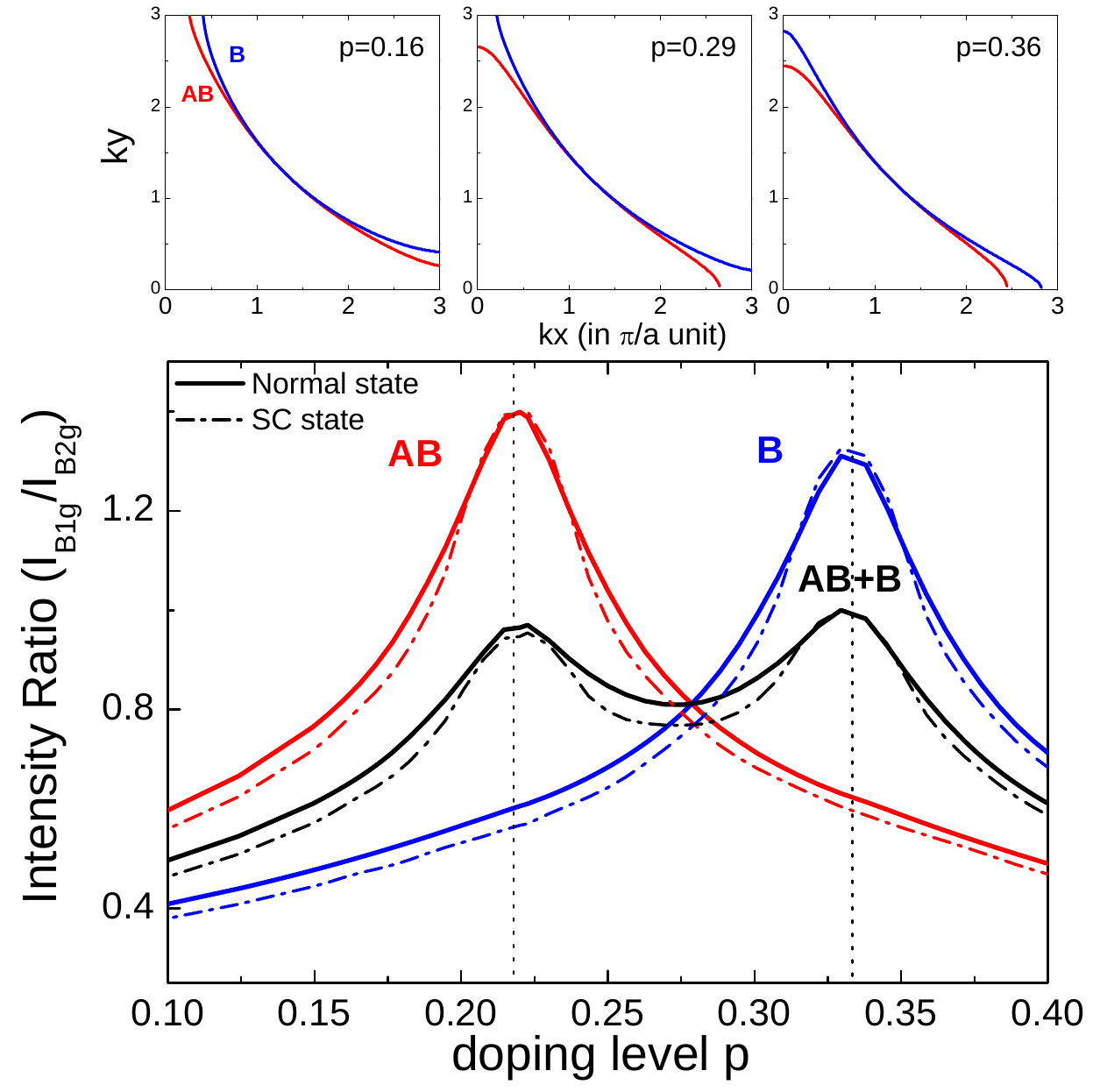}
\end{center}
\caption{Integrated intensity ratios $I_{B1g}/I_{B2g}$ for the normal (solid lines) and
superconducting (dash-dot lines) phases over a doping range wider than what is accessible experimentally.
The contributions from the AB (red/light gray) and B (blue/dark gray) bands are shown separately. They peak
at $p=0.22$ and
0.33 respectively, which correspond to the Lifshitz transitions in each of these bands, as shown
in the top panel. Note that around  $p=0.22$ the AB band contribution is dominant.
The physical intensity ratios obtained from the total responses, i.e., the sum of the
responses from the two bands, is shown in black. All the curves for a given phase are normalized by the value
of the total response in that phase at $p=0.22$.}
\label{fig9}
\end{figure}

In this doping range the B band is far from its own Lifshitz transition, which would take place at a much higher doping
$p=0.33$ which is inaccessible experimentally.
Therefore, its contribution to the Raman \BAN response (blue/ dark gray) curves in Fig. \ref{fig8}) is less relevant
than the AB band one (red/light gray) around $p = 0.22$.

For the sake of completeness in Fig.~\ref{fig9} we show the calculated ratio of the total integrated Raman intensities
$I_{\BAN}/I_{\BN}$ (defined in main text)
for a doping range larger than what is accessible experimentally.
Two peaks are clearly discernable, both in the normal and the superconducting cases, at $p = 0.22$ and at $p=0.33$,
the first corresponding to the AB band Lifshitz transition, the second to that of the B band.

For clarity we also plot the $I_{\BAN}/I_{\BN}$ for the AB and B bands separately, which shows
that the contribution of each band peaks at its respective Lifshitz transition point.

In particular, as we previously stated, the B band
contribution to the Raman response at $p = 0.22$ doping is substantially smaller than the AB band one (while the converse
is true at $p = 0.33$). We also display in the top panels of  Fig.~\ref{fig9} the Fermi surface in the first quadrant of the Brillouin zone. We clearly observe the change of topology from an hole- to electron-like Fermi surface in the AB and B bands, for $p = 0.22$ and for $p=0.33$ respectively.

\clearpage

\bibliography{biblioarxiv}

\begin{thebibliography}{59}
\bibitem[{not()}]{note1}
\bibinfo{note}{The phonon peak contribution is less than 3\% of the total
  integrated area}.
\end{thebibliography}

\end{document}